\documentclass[%
 reprint,
 superscriptaddress,
 amsmath,amssymb,
 aps,
 pre,
]{revtex4-1}
\usepackage{graphicx}
\usepackage{dcolumn}
\usepackage{bm}
\usepackage{upgreek}
\usepackage{bm}
\usepackage{amsmath}
\usepackage{color}
\begin{document}
\preprint{APS/123-QED}

\title{Sticking without contact: Elastohydrodynamic adhesion}
\author{Vincent Bertin}
\affiliation{Physics of Fluids Group, University of Twente, 7500 AE Enschede, The Netherlands}
\affiliation{Aix Marseille Univ, CNRS, IUSTI UMR 7343, Marseille 13453, France}

\author{Alexandros T. Oratis}
\affiliation{Physics of Fluids Group, University of Twente, 7500 AE Enschede, The Netherlands}

\author{Jacco H. Snoeijer}
\affiliation{Physics of Fluids Group, University of Twente, 7500 AE Enschede, The Netherlands}

\date{\today}

\begin{abstract}
The adhesion between dry solid surfaces is typically governed by contact forces, involving surface forces and elasticity. For surfaces immersed in a fluid, out-of-contact adhesion arises due to the viscous resistance to the opening of the liquid gap. While the adhesion between dry solids is described by the classical JKR theory, there is no equivalent framework for the wet adhesion of soft solids. Here, we investigate theoretically the viscous adhesion emerging during the separation of a sphere from an elastic substrate. The suction pressure within the thin viscous film between the solids induces significant elastic displacements. Unexpectedly, the elastic substrate closely follows the motion of the sphere, leading to a sticking without contact. The initial dynamics is described using similarity solutions, resulting in a nonlinear adhesion force that grows in time as $F \propto t^{2/3}$. When elastic displacements become large enough, another similarity solution emerges that leads to a violent snap-off of the adhesive contact through a finite-time singularity. The observed phenomenology bears a strong resemblance with JKR theory, and is relevant for a wide range of applications involving viscous adhesion.   

\end{abstract}

\maketitle

\section{Introduction}

Adhesion between particles is a central problem in soft matter physics, with relevance across fields like cohesive materials (e.g., colloidal gels)~\cite{lu2013colloidal}, cellular biophysics~\cite{chu2005johnson,maitre2012adhesion}, and industrial applications such as pressure-sensitive adhesives~\cite{creton2003pressure}. The softness is often crucial, as soft solids can conform to rough surfaces, enhancing adhesion through intimate contact. Over the past 50 years, fundamentals models like Johnson–Kendall–Roberts (JKR)~\cite{johnson1971surface} and Derjaguin-Muller-Toporov (DMT)~\cite{derjaguin1975effect} have shaped our understanding of adhesion. More recent works have extended this knowledge by accounting for surface roughness~\cite{persson2001effect}, and the critical role of surface stresses at small scales~\cite{style2013surface,salez2013adhesion,karpitschka2016surface,jensen2017strain,liu2019effects}.

An important aspect of adhesion is its dynamics—specifically, how the speed of separation influences adhesive strength. The viscoelastic nature of soft solids introduces complexities, such as a rate-dependent adhesive strength~\cite{creton2016fracture}, fibrillation or the formation of fingering instabilities~\cite{amar2005fingering,nase2008pattern,karnal2023interface}. The relaxation post-snapping of an adhesive can also be influenced by the poroelastic properties of the solid~\cite{berman2019singular,xu2020viscoelastic}. In many applications, dynamical adhesion is mediated by a thin viscous layer without any direct solid-solid contact. A popular example is the capture of prey by chameleons or frogs, which relies on the viscous nature of their saliva~\cite{brau2016dynamics,noel2017frogs}. The standard framework of this viscous adhesion, often referred to as Stefan adhesion~\cite{stefan1875versuche}, considers two rigid solids immersed inside a viscous fluid. When pulled apart, an adhesive force arises from the fluid resistance to the opening of the thin liquid gap \cite{bikerman1947fundamentals}. For rounded solids with an effective curvature $1/R$, this force is given by the Reynolds force 
\begin{equation}
\label{eq:StefanForce}
F = \frac{6\pi \eta R^2 V}{\delta_0},
\end{equation} 
where $\eta$, $V$, and $\delta_0$, respectively, are the liquid viscosity, separation speed, and liquid gap~\cite{leal2007advanced}. 

\begin{figure}
	\centerline{\includegraphics{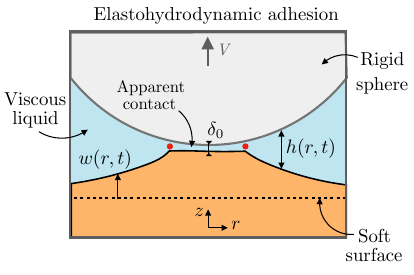}}
	\caption{Schematic of the elastohydrodynamic adhesion problem. A rigid sphere of radius $R$ is pulled-up at a speed $V$ from a soft surface. We represent here a magnified picture near the contact region. The initial sphere-surface distance is denoted $\delta_0$, the elastic displacement field of the surface $w(r,t)$, and fluid thickness $h(r,t)$. The red dots indicate the location of the apparent contact (see Sec.~\ref{sec:snapping}).}
	\label{fig:fig1}
\end{figure}	 

While of fundamental importance, the classical equation (\ref{eq:StefanForce}) completely fails to describe the viscous adhesion between very soft solids. Elastic deformations are central to the JKR theory of dry adhesion~\cite{johnson1971surface}, and are naturally expected to be significant for viscous adhesion (e.g. for prey capture). Indeed, recent experiments have shown that switching from rigid to soft surfaces fundamentally changes the nature of viscous adhesion~\cite{shao2023out}. The principle of this so-called \textit{elastohydrodynamic adhesion}~\cite{wang2020dynamic} is illustrated in Fig.~\ref{fig:fig1} for the geometry of a sphere pulled away from a soft solid. The viscous suction pressure induces elastic displacements that qualitatively change the nature of the retraction flow. Yet, a detailed understanding of the adhesive force and the detachment of the contact is currently lacking.

In this paper we theoretically describe the principles of elastohydrodynamic adhesion. We show that the dynamics can be decomposed into two phases. During the initial phase, the sphere seems to stick to the surface, as sketched in Fig.~\ref{fig:fig1}. The adhesive force increases via a non-trivial power law 
\begin{equation}
\label{eq:StickingForce}
F(t) =  -5.94\, E^{*2/3}\eta^{1/3} RV t^{2/3},
\end{equation}
where $E^*$ is the reduced Young's modulus. This result replaces the classical formula (\ref{eq:StefanForce}). Remarkably, the initial force is completely independent of the initial separation, and regularises the divergence near contact implied in (\ref{eq:StefanForce}). However, the scaling $F \sim t^{2/3}$ implies a large effective stiffness at early times, in line with experimental observation~\cite{shao2023out}. The second phase of the dynamics leads to the snapping of the soft surface, that we identify as a finite-time singularity. As such, the phenomenology of soft elastohydrodynamic adhesion bears a strong qualitative resemblance with the JKR theory of dry contacts, and we highlight similarities and differences. 

The paper is organized as follows. The elastohydrodynamic model is introduced in Sec.~\ref{sec:EHD-adhesion}. The separation dynamics is then analyzed by numerical solutions in Sec.~\ref{sec:dynamics}, highlighting the rigid and soft limits. Then, the sticking and snapping regimes are described analytically in terms of similarity solutions Sec.~\ref{sec:asymptotics}, by which we reveal the fine-structure of the pressure and deformations inside the viscous layer.  The paper closes with an outlook and a comparison with JKR adhesion in Sec.~\ref{sec:conclusion}.

\section{Elastohydrodynamic adhesion}
\subsection{Model}
\label{sec:EHD-adhesion}

In this work, we employ an elastohydrodynamic model based on the theory of soft lubrication~\cite{rallabandi2024fluid}. This framework has gained significant attention in recent years and has been successfully applied to various contexts, including friction forces~\cite{saintyves2016self,dong2023transition}, film drainage~\cite{wang2015out}, fingering instability~\cite{pihler2012suppression,lister2013viscous,peng2020viscous}, capillary flows~\cite{rivetti2017elastocapillary,bertin2022enhanced,tamim2023spreading} and contactless rheological measurements~\cite{leroy2011hydrodynamic,leroy2012hydrodynamic,zhang2022contactless}. 

We consider a rigid sphere, submerged in a viscous fluid. The sphere is initially placed at a distance $\delta_0$ from a soft planar surface and suddenly pulled upward at $t=0$ with a vertical speed $V$. The system is axisymmetric and the radial position is denoted by $r$. We assume the sphere-plane distance to be much smaller than the sphere radius and describe the flow in the liquid gap with the lubrication approximation. The liquid film thickness $h(r,t)$ (see Fig.~\ref{fig:fig1}) satisfies the thin-film equation~\cite{oron1997long}
\begin{equation}
	\label{eq:thin-film}
	\frac{\partial h(r,t)}{\partial t} = \frac{1}{12\eta r}\frac{\partial }{\partial r} \left(r h^3(r,t) \frac{\partial p(r,t)}{\partial r}\right),
\end{equation}
where $p(r,t)$ is the fluid pressure field. In the flow region, the rigid sphere surface can be modeled with a parabolic approximation such that the liquid film thickness is written as~\cite{rallabandi2024fluid}
\begin{equation}
	\label{eq:film_thickness}
	h(r,t) = \delta_0 + Vt + \frac{r^2}{2R} - w(r,t),
\end{equation}
where $w(r,t)$ represents the elastic displacement field of the surface (see Fig.~\ref{fig:fig1} (b)). The latter is modeled using linear elasticity theory and is related to the pressure field via a convolution integral involving the elastic Green's function $\mathcal{M}$. Treating the substrate as a semi-infinite medium, the elastic displacement reads~\cite{davis1986elastohydrodynamic,johnson1987contact}
\begin{align}
	\label{eq:elasticity}
    w(r,t) &= -\frac{4}{\pi E^*}\int_0^\infty \mathcal{M}(r,x) p(x,t) \, \mathrm{d}x, \\
    \mathcal{M}(r,x) &= \frac{x}{r+x}K\left(\frac{4rx}{(r+x)^2}\right).
\end{align}
The function $K$ denotes the complete elliptic integral of the first kind. The system of equations~\eqref{eq:thin-film}-\eqref{eq:elasticity} is similar to those derived by Davis et al. who studied the collision between spheres in a liquid~\cite{davis1986elastohydrodynamic}. Here, we focus on a very different regime: a near-contact initial condition where the sphere is withdrawn at a constant speed. We solve \eqref{eq:thin-film}-\eqref{eq:elasticity} numerically using a finite difference scheme similar to the Refs. \cite{liu2022lubricated,bertin2024similarity}. An adaptive mesh refinement is used to resolve the singularity near the apparent contact radius that is discussed later. 

To highlight the non-trivial scaling laws that emerge, we prefer not to introduce dimensionless variables, but rather present results using the relevant scales. The lubrication pressure scales as $\eta R V/h^{*2}$, where $h^*$ is a typical film thickness, while the Hertz-like elastic pressure scales as $E^* h^*/\sqrt{Rh^*}$. Balancing these pressures allows to find the characteristic elastohydrodynamic film thickness $h^*$ together with the characteristic pressure $p^*$ and force $F^*$, as
\begin{equation}
\label{eq:EHD_scales}
\begin{split}
    h^* = R \left(\frac{\eta V}{E^* R}\right)^{2/5}&, \quad \quad p^* = \frac{\eta V}{R} \left(\frac{\eta V}{E^* R}\right)^{-4/5}, \\
    F^* &= \eta R V \left(\frac{\eta V}{E^* R}\right)^{-2/5}.
\end{split}
\end{equation}
The comparison of the elastohydrodynamic film thickness with the initial gap thickness $\delta_0$ defines the only dimensionless number of the problem, as
\begin{equation}
\frac{h^*}{\delta_0} = \left(\frac{\eta V R^{3/2}}{E^* \delta_0^{5/2}}\right)^{2/5}.
\end{equation}
The dimensionless number $h^*/\delta_0$ compares the viscous lubrication forces for a liquid gap $\delta_0$ to elastic forces. The variation of the parameter $h^*/\delta_0$ can be interpreted as a change of initial gap if all other parameters are fixed. Equivalently, the parameter $h^*/\delta_0$ is a measure of the softness of the substrate, as it is larger for smaller elastic moduli.

\begin{figure*}
	\centerline{\includegraphics{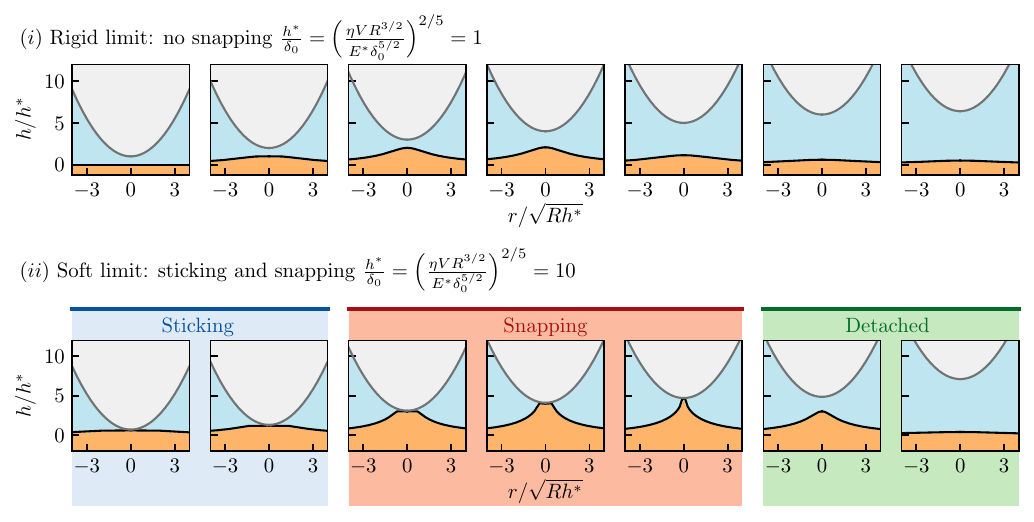}}
	\caption{Series of snapshots showing the detachment dynamics. $(i)$ Typical example of the rigid limit (small to moderate $h^*/\delta_0$, here set to $1$). $(ii)$ Typical example of the soft limit for large  (large $h^*/\delta_0$, here set to $10$). The dimensionless times are respectively $t/(h^*/V) = 0,\, 1,\, 2,\, 3, \, 4,\, 5,\, 5.4$ in $(i)$ and $t/(h^*/V) = 0.6, \, 1.2,\, 3.0,\, 4.0,\, 4.6,\, 4.75,\, 7.0$ in $(ii)$. In the soft limit, one observes a ``sticking" phase and a ``snapping" phase, as indicated by blue and red background shades.}
	\label{fig:fig2}
\end{figure*}

\subsection{Dynamics}
\label{sec:dynamics}
The separation dynamics is illustrated in Fig.~\ref{fig:fig2} with snapshots of the elastic displacements for two values of the $h^*/\delta_0$, which correspond to Supp. Video 1 and 2 respectively. The separation clearly exhibits two regimes depending on the softness: $(i)$ a ``rigid" regime where the elastic deformations remain relatively small, and $(ii)$ a ``soft" regime that is characterized by large deformations, and which gives rise to sticking and snapping: even though the sphere and substrate are separated by a thin liquid layer, the soft surface seems to stick to the sphere until the apparent contact violently snaps back and returns to its equilibrium configuration. We will show that this soft regime is highly intricate, yet the most relevant for applications, and is therefore the focus of this manuscript.

\begin{figure*}
	\centerline{\includegraphics{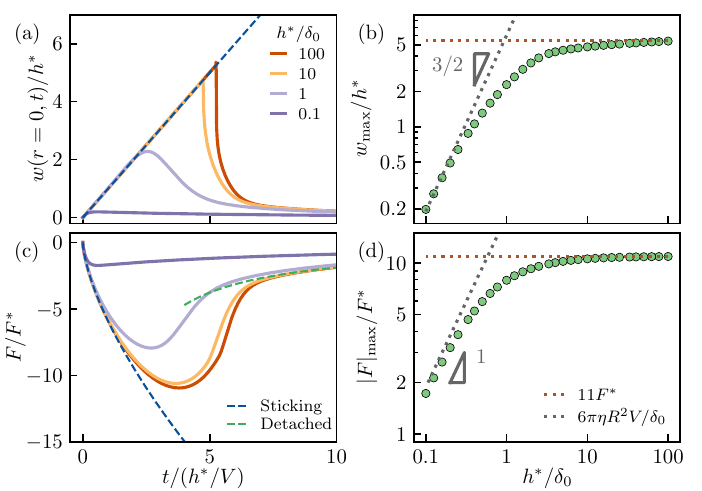}}
	\caption{ Rescaled central elastic displacement (a) and dimensionless force (c) versus rescaled time for four different values of the dimensionless elastic parameter $h^*/\delta_0$. The blue dashed lines show the sphere displacement $Vt$ in (a) and the sticking similarity solution Eq.~\eqref{eq:StickingForce} in (c). The green dashed line in (c) displays the detached lubrication force Eq.~\eqref{eq:rigid} for $\delta_0=0$. Maximum central elastic displacement (b) and adhesion force as a function of $h^*/\delta_0$. The gray dotted lines exhibit a scaling law $(h^*/\delta_0)^{-3/2}$ in (b) and the rigid limit $6\pi\eta R^2V/\delta_0$ in (d). The brown dotted line shows constant values $w_\mathrm{max} = 5.4\, h^*$ and $\vert F\vert_\mathrm{max} = 11 F^*$. }
	\label{fig:fig3}
\end{figure*}

The dynamics are further quantified in Fig.~\ref{fig:fig3}(a) showing the central elastic displacement $w(0,t)$ versus time shown for different values of $h^*/\delta_0$. The central elastic displacement first increases linearly in time as $w = Vt$ (see blue dashed line in Fig.~\ref{fig:fig3}(a)), then reaches a maximal value $w_\mathrm{max}$ before decaying at long times. Figure~\ref{fig:fig3}(c) reports the corresponding adhesion force $F(t)$, which is obtained by integrating the pressure in the film. The force increases very steeply at early times, and passes through a maximum value $F_{\rm max}$ around the same instant that the maximum displacement $w_\mathrm{max}$. At very late times the elastic displacement goes to zero and the adhesion force follows the lubrication force for undeformed detached surfaces  
\begin{equation}
\label{eq:rigid}
F(t) = -\frac{6\pi \eta R^2 V}{(\delta_0 + Vt)},
\end{equation}
as shown in green dashed lines in Fig.~\ref{fig:fig3}(c). This expression corresponds to the classical Reynolds force Eq.~\eqref{eq:StefanForce} for a gap distance $\delta_0 + Vt$, often used in viscous adhesion. However, as is clear from the figure, this classical formula does not at all capture the most interesting part of the elastohydrodynamic adhesion.

The dynamics for small $h^*/\delta_0$ (rigid limit $(i)$) are relatively easy to understand, since the elastic displacements are small compared to the liquid gap at all times. The elastic displacement field remains smooth throughout the full dynamics (see Supp. Video 1 and purple curves in Fig.~\ref{fig:fig3}(a)). One can determine the typical scale of the elastic displacements in this regime from the elasticity equation as $w_\mathrm{rig} \sim p_\mathrm{rig} r_\mathrm{rig}/E^*$, where $p_\mathrm{rig} = \eta V R/\delta_0^2$ and $r_\mathrm{rig} = \sqrt{R\delta_0}$ are the typical hydrodynamic lubrication pressure and radial scales for rigid surfaces, which leads to elastic displacement $\sim \eta V R^{3/2}/(E^*\delta_0^{3/2})$~\cite{bertin2022soft}. Indeed, the maximum elastic displacement follows this scaling law at small $h^*/\delta_0$ (see dotted gray lines in Fig.~\ref{fig:fig3}(b)). The corresponding maximum adhesion is given by the Reynolds lubrication force~\eqref{eq:rigid}, as $6\pi\eta R^2 V/\delta_0$ (see dotted gray lines in Fig.~\ref{fig:fig3}(d)). 

The interest of the paper, however, lies in the regime where $h^*/\delta_0$ is large (soft limit, cf. Fig.~\ref{fig:fig2}$(ii)$ and Supp. Video 2). In that limit, the elastic displacements largely exceed the liquid film thickness. The maximum elastic displacement is typically $w_\mathrm{max }\approx 5 \, h^*$ and does not depend on the initial gap thickness in the $h^*/\delta_0 \gg 1$ limit (see brown dotted line in Fig.~\ref{fig:fig3}(b)). Similarly, the maximum adhesive force is independent of the initial liquid gap $\delta_0$ in this limit and is typically $\vert F\vert_\mathrm{max} \approx 11 \, F^*$. These results for the maximum force and maximum displacement confirm the validity of scalings of equation~\eqref{eq:EHD_scales}, in the soft limit. 

In what follows, we will show that for soft elastohydrodynamic adhesion one can identify two distinct phases before snapping, which we refer to as “sticking phase" and “snapping phase". These are, respectively, indicated by the blue and red shades in Fig. \ref{fig:fig2}$(ii)$. Indeed, until the elastic displacement reaches its maximum value, the liquid gap near the center of the sphere does not change: as if the soft surface effectively sticks to the sphere. At a certain time $t_c$, the elastic surface snaps and the elastic displacement decays very rapidly on a time scale smaller than $h^*/V$. 
The remainder of the article is dedicated to revealing the physical nature of sticking and snapping.

\begin{figure*}
	\centerline{\includegraphics{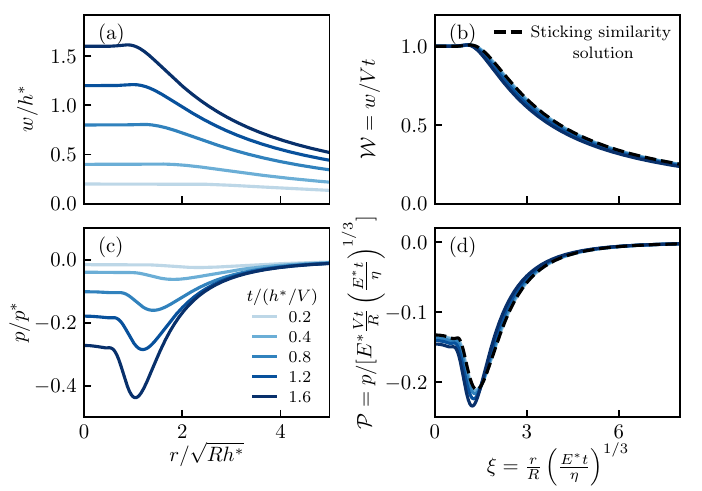}}
	\caption{Dimensionless elastic displacement (a) and pressure (c) as a function of the dimensionless radius for five different times (resp. $t = 0.2, 0.4, \, 0.8, \, 1.2$ and $1.6 \, h^*/V$) during the sticking phase. The ratio $h^*/\delta_0$ is set to $10$, corresponding to the snapshots in Fig.~\ref{fig:fig2}$(\mathrm{b}_{ii})$. In (b) and (d), the profiles are rescaled by the typical length and pressure scales of the sticking phase, corresponding to Eq. \eqref{eq:elastic_phase_scale}. The similarity solutions of Eqs.~\eqref{eq:thin-film_self-similar}-\eqref{eq:elasticity_self_similar} are shown in black dashed lines.}
	\label{fig:fig4}
\end{figure*}	 

\section{Asymptotic analysis}
\label{sec:asymptotics}
\subsection{Early-times: sticking similarity solution} 
\label{sec:sticking}

We first address the early-time dynamics of the adhesive force, through a quantitative analysis of the sticking phase. Figures~\ref{fig:fig4}(a) and (c) report the elastic displacement and pressure profiles at different times. Since the profiles at different times exhibit similar shapes, we seek for self-similar solutions of the displacement and pressure fields. For this we wish to identify the relevant scales in the sticking regime, which we will denote with the subscript ‘‘st''. The liquid gap initially does not drain and the elastic displacement grows following the sphere displacement, such that $w_\mathrm{st} = Vt$, as shown in Fig.~\ref{fig:fig3}(a). The elasticity equation \eqref{eq:elasticity} prescribes that the elastic displacement and pressure scales are related via $w_\mathrm{st} = p_\mathrm{st}r_\mathrm{st}/E^*$. At early times, the radial scale $r_\mathrm{st}$ will turn out to be large such that that the soft surface curvature $w_\mathrm{st}/r_\mathrm{st}^2$ is small compared to the sphere curvature. Therefore, the film thickness is set by the curvature of the sphere as $h_\mathrm{st} = r^2_\mathrm{st}/R$. Injecting the latter relation in the thin-film equation leads to $\eta V = h^3_\mathrm{st} p_\mathrm{st}/r_\mathrm{st}^2 = r^4_\mathrm{st}p_\mathrm{st}/R^3$. Combining the aforementioned expressions leads to the scales during the sticking phase,
\begin{equation}
\label{eq:elastic_phase_scale}
\begin{split}
w_\mathrm{st} &= Vt, \quad \quad p_\mathrm{st} = E^* \frac{Vt}{R} \left(\frac{E^*t}{\eta}\right)^{1/3},  \\ & r_\mathrm{st} = R \left(\frac{\eta}{E^*t}\right)^{1/3}. 
\end{split}
\end{equation}
The typical radial scale found here decreases with time as $t^{-1/3}$, while the pressure increases as $t^{4/3}$. To test whether these arguments correctly capture the dynamics, we rescale the displacement and pressure profiles according to (\ref{eq:elastic_phase_scale}): Figures~\ref{fig:fig4}(b) and (d) indeed provide an excellent collapse of the data.

We now reveal the detailed structure of the sticking phase by means of a similarity analysis. Using the scales \eqref{eq:elastic_phase_scale}, we introduce the self-similar Ansatz 
\begin{equation}
\label{eq:elastic-similarity-ansatz}
\begin{split}
w(r,t) = Vt \, &\mathcal{W}(\xi), \quad  p(r,t) = E^* \frac{Vt}{R} \left(\frac{E^*t}{\eta}\right)^{1/3} \mathcal{P}\left(\xi\right), \\ & \text{with}\quad \xi = \frac{r}{R} \left(\frac{E^*t}{\eta}\right)^{1/3}.
\end{split}
\end{equation}
We inject Eq.~\eqref{eq:elastic-similarity-ansatz} into Eq.~\eqref{eq:thin-film}-\eqref{eq:film_thickness} which at early times gives a time-independent equation
\begin{equation}
\label{eq:thin-film_self-similar}
1 - \mathcal{W} - \frac{\xi \mathcal{W}'}{3} = \frac{1}{12 \xi} \left(\frac{\xi^7 \mathcal{P}'}{8} \right)',
\end{equation}
where the primes denote derivative with respect to the similarity variable $\xi$. Likewise, the elasticity equation \eqref{eq:elasticity} can be rewritten as
\begin{equation}
	\label{eq:elasticity_self_similar}
	\mathcal{W}(\xi) = -\frac{4}{\pi}\int_0^\infty \mathcal{M}(\xi,\tilde{\xi}) \mathcal{P}(\tilde{\xi}) \, \mathrm{d}\tilde{\xi}, 
\end{equation}
The coupled set of equations \eqref{eq:thin-film_self-similar}-\eqref{eq:elasticity_self_similar} is free from parameters and has a universal solution with boundary conditions $\mathcal{P}'(0) = 0$ and $\mathcal{P}(\infty) = 0$, corresponding the the symmetry and far-field condition respectively. The similarity solutions obtained from solving ~\eqref{eq:thin-film_self-similar}-\eqref{eq:elasticity_self_similar} are superimposed as the black dashed lines in Fig. \ref{fig:fig4}(b) and (d). The similarity solutions offer an perfect description of the numerical data during the sticking phase, without any adjustable parameters.

The similarity solutions can be leveraged to obtain the nonlinear adhesive force, Eq.~\eqref{eq:StickingForce}, shown as the blue dashed in see Fig.~\ref{fig:fig3}(c). Namely, the force can be computed from the similarity solutions as 
\begin{eqnarray}
\label{eq:elastic-force}
F(t) &=& 2\pi\int_{\mathbb{R}_+} p(r,t) r\mathrm{d}r 
= 2\pi\int_{\mathbb{R}_+} p_\mathrm{st}  r_\mathrm{st}^2 \, \mathcal P(\xi) \xi \mathrm{d}\xi \nonumber \\
&=& -C\, E^* RVt \left(\frac{\eta}{E^* t}\right)^{1/3}.
\end{eqnarray}
The scalings follow directly from Eq.~\eqref{eq:elastic_phase_scale}, while the similarity solution $\mathcal P(\xi)$ can be used to compute the prefactor 
\begin{equation}
C = -2\pi \int_{\mathbb{R}_+} \mathcal{P}(\xi)\, \xi \mathrm{d}\xi \approx 5.94,
\end{equation}
such that we recover Eq.~\eqref{eq:StickingForce}. The prediction of the adhesion force is again in excellent agreement with the full numerical solution at early times as shown in Fig.~\ref{fig:fig3}(c) (see blue dashed lines). We emphasise that the early-time adhesive force is not linear in time and depends on the viscosity. Hence, the soft surface cannot be modeled as an effective spring of stiffness $k \sim E^*\sqrt{R\delta_0}$ at early times, as was suggested in Ref.~\cite{shao2023out}.

\subsection{Snap-off is a finite-time singularity}  
\label{sec:snapping}
\begin{figure*}
	\centerline{\includegraphics{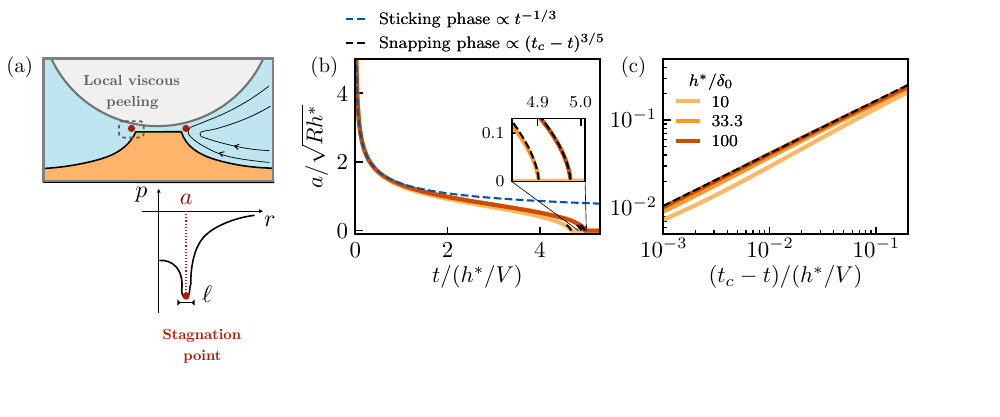}}
	\caption{(a) Schematic of the streamlines of the flow before the snapping as well as the pressure field. The minimal pressure is off-centred (see bottom schematic) such that the red dot is a stagnation point of the flow. The location of the stagnation point defines the apparent contact radius $a$, while $\ell$ denotes the typical radial extent of the pressure spike. (b) Dimensionless apparent contact radius versus the dimensionless time, where different colors correspond to varying $h^*/\delta_0$. The blue dashed lines display the scaling law from the sticking phase (see Eq.~\eqref{eq:sticking-contact_radius}), which works well at early times, but does not capture the finite-time snap-off. The inset shows a zoom-in the near snap-off region. (c) The same data as in (b) are plotted in log-log using the time-to-snapping $t_c - t$ as a variable, revealing the scaling relation Eq.~\eqref{eq:snapping-contact_radius}. }
	\label{fig:fig5}
\end{figure*}	 

The soft surface exhibits a clear snap-off at a given time, which obviously cannot be described by the similarity analysis of Sec.~\ref{sec:sticking}. Indeed, the adhesive force starts to deviate from the $t^{2/3}$ law at a time  around $2 h^*/V$ (Fig.~\ref{fig:fig3}(c)). This departure preceeds a phase that ultimately leads to detachment, referred to as the snapping phase. Examining the surface displacement profiles in this phase (red shades in Fig.~\ref{fig:fig2}($ii$)), we observe that the soft surface remains flat at the center over a certain radial extent. We interpret this radial extent as the apparent contact area. A strong curvature of the surface develops at the edge of this zone. 

We recall here that there is no true contact and that a lubrication film separates the sphere from the surface surface. Before proceeding, we thus need to carefully define what is meant by ``contact". A noticeable feature of the pressure in Fig.~\ref{fig:fig4}(c)-(d) is that it has an off-centered global minimum at a position $r=a$, which has significant consequences for the flow field. In viscous flows, the location of a pressure minimum is a stagnation point of the flow. Hence, the streamlines of the flow far from the center do not enter the region $r<a$ (see Fig.~\ref{fig:fig5}(a)), as already pointed out in Ref.~\cite{shao2023out}. This feature qualitatively explains why the soft surface appears to stick to the sphere without direct contact in Fig.~\ref{fig:fig2}($ii$): at a given time $t$, no fluid escaped the region $r<a(t)$. We therefore call $a$ the \textit{apparent contact radius}. The detachment of the apparent contact is not driven by a flow out of the contact, but by a local viscous ``peeling" around that is described below. 

\begin{figure*}
	\centerline{\includegraphics{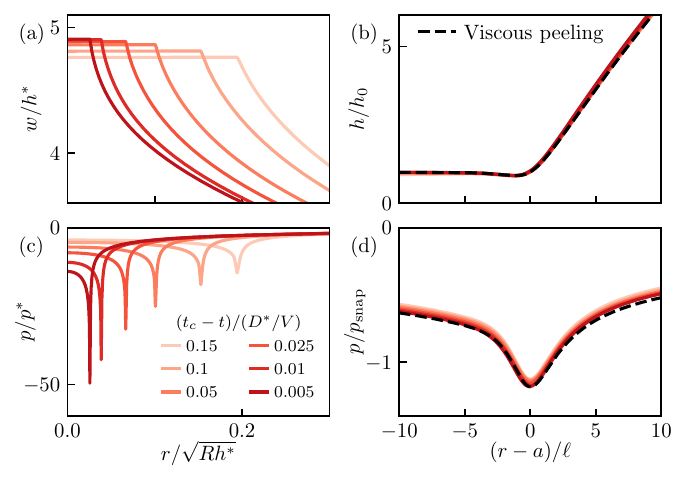}}
	\caption{Dimensionless elastic displacement (a) and pressure (c) as a function of the dimensionless radius for six different time-to-snapping near $t_c$. The ratio $h^*/\delta_0$ is set to $100$. In (b) and (d), the profiles are rescaled by the typical length and pressure scales of the pressure spike, corresponding to Eq. \eqref{eq:snapping_scales}. The similarity solutions of Eqs.~\eqref{eq:LLD-like}-\eqref{eq:Hilbert} are shown in black dashed lines.}
	\label{fig:fig6}
\end{figure*}	 

To understand the detachment of the contact and the snapping characteristics, we analyze the time evolution of the apparent contact radius plotted in Fig.~\ref{fig:fig5}(b). The apparent contact radius decreases monotonically in time before reaching zero at a finite time $t_c$, corresponding to the snapping time. At early times, $a$ is very well approximated by the prediction of the sticking similarity solution 
\begin{equation}
\label{eq:sticking-contact_radius}
a(t) =  A_0 \, R(\eta t/E^*)^{-1/3},
\end{equation}
which is plotted as the dashed blue line in Fig.~\ref{fig:fig5}(b). The prefactor $A_0 \approx 1.37$ corresponds to the $\xi$-value at the minimum of $\mathcal{P}$ (see Fig.~\ref{fig:fig4}(d)). However, significant deviations from the sticking similarity solutions arise at later times (typically $t > 2h^*/V$), around which where the curve $a(t)$ gradually turns concave. The inset of Fig.~\ref{fig:fig5}(b) shows that the contact radius $a \to 0$ at a well-defined critical time $t_c$. Replotting the contact radius versus $t_c-t$ on double logarithmic axis in Fig.~\ref{fig:fig5}(c), we see the emergence of a scaling regime, where $a$ follows a power law $(t_c-t)^n$ with $n < 1$. The exponent seems to depend slightly on the value of $h^*/\delta_0$, where a value close to $3/5$ is is found for large $h^*/\delta_0$. The process is accurately described by an empirical fit
\begin{equation}
\label{eq:snapping-contact_radius}
a(t) = A_1\sqrt{Rh^*} \left(\frac{t_c-t}{h^*/V}\right)^{3/5},
\end{equation}
with $A_1 \approx 0.66$. Hence, the radial velocity $\dot{a}$ blows up algebraically near snapping. Such a behavior is reminiscent of finite-time singularities that are often found in numerous dynamical systems governed by non-linear partial differential equations~\cite{eggers2015singularities}. A prime example is the drop pinch-off process, where the neck radius near the break-up follows dynamics analogous to the snap-off described here~\cite{eggers1997nonlinear}.

\subsection{Late times: peeling similarity solution}

We now show how the snapping proceeds via a peeling-like motion that, once again,  can be described by a similarity solution. The elastic displacement and pressure fields at various time-to-snapping are shown in Fig.~\ref{fig:fig6}(a)-(c). Similar to the radial velocity, the pressure field near the center axis also diverges as the time-to-snapping $t_c-t$ goes to zero. The pressure field adopts a localized spike at a given radius that corresponds to the edge of the apparent contact area (see Fig.~\ref{fig:fig6}(a) and (c)). To describe analytically the singular features of the snapping, we introduce the typical length scale $\ell$ of the pressure spike (see Fig.~\ref{fig:fig5}(a)) and the contact radius retraction speed $u = -\dot{a}$. The pressure must be regularized by lubrication effects leading to the scaling relation $p/\ell \sim \eta u/h_a^2$, where $h_a$ is the local film thickness at the edge of the apparent contact area. As the liquid film within the apparent contact area does not fill in, $h_a$ is set by the initial condition, which gives here $h_a = \delta_0 + a^2/(2R)$. Furthermore, we neglect the parabolic term due to the sphere curvature in \eqref{eq:film_thickness} such that the film thickness is approximately $h \simeq h_a + Vt - w(r,t)$. Hence, the elastic strain is equal to the gradient of the film thickness as $w'(r) \simeq - h'(r) \sim h_a/\ell$. Within linear elasticity, the elastic stress is proportional to strain, which gives the scaling relation $p \sim E^* h_a/\ell$. The latter relation, combined with the lubrication scaling, allows to identify the scales of the pressure spike, denoted $p_\mathrm{snap}$, as 
\begin{equation}
\label{eq:snapping_scales}
p_\mathrm{snap} = \left(\frac{6 E^*\eta u}{h_a}\right)^{1/2}, \quad \quad \ell = h_a \left(\frac{E^*h_a}{24 \eta u}\right)^{1/2},
\end{equation}
where the prefactors are chosen for later convenience. As shown in Fig.~\ref{fig:fig6}(b)-(d), the film thickness and pressure indeed collapse when plotted with the scales~\eqref{eq:snapping_scales}, which validates our approach. More formally, we introduce the following similarity Ansatz
\begin{equation}
\label{eq:snapping-similarity-ansatz}
\begin{split}
h(r,t) = h_a \, H&(X), \quad  p(r,t) = p_\mathrm{snap}P\left(X\right), \\ & X = \frac{r-a(t)}{\ell}.
\end{split}
\end{equation}
Assuming the scale separation $h_a \ll \ell \ll a \ll R$ and injecting the Ansatz in the thin-film equation yields 
\begin{equation}
H' = \left(H^3 P'\right)'.
\end{equation}
The equation can be integrated once to obtain a Landau-Levich type of equation~\cite{landau1942dragging}
\begin{equation}
\label{eq:LLD-like}
P' = \frac{H-1}{H^3},
\end{equation}
where the integration constant has been set to unity from the matching condition to the local film thickness at the edge of the contact $h(r\to a^-) = h_a = h_a H(X\to -\infty)$. A very similar treatment has been performed to describe the peeling of elastic blisters~\cite{lister2013viscous,peng2020viscous}, the lubricated Hertzian contact~\cite{snoeijer2013similarity}, and the bouncing of a sphere on an elastic surface~\cite{wang2013central,venner2016central,bertin2024similarity}. The length scale of the pressure spike being much smaller than the apparent contact radius $\ell \ll a$, the elasticity in the region can be treated as quasi 2d. In other words, the elastic kernel $\mathcal{M}$ simplifies to the 2d semi-infinite elastic solution in the boundary layer. As shown in \cite{snoeijer2013similarity,bertin2024similarity}, this leads to the integral equation
\begin{equation}
\label{eq:Hilbert}
H''(X) = \frac{1}{\pi} \, \mathrm{p.v.}\int_\mathbb{R} \frac{P'(\bar{X})}{X-\bar{X}}\, \mathrm{d}\bar{X},
\end{equation}
where we recover the Hilbert transform. The integral in \eqref{eq:Hilbert} should be understood as the Cauchy principal value ($\mathrm{p.v.}$). Interestingly, the set of equations \eqref{eq:LLD-like} and \eqref{eq:Hilbert} are identical to those describing the outlet region of the lubricated Hertzian contact~\cite{snoeijer2013similarity}. However, a crucial difference with respect to the Hertzian contact problem concerns the boundary conditions, which for the Hertzian profile dictates an asymptotic matching condition $H(X\to \infty) \propto X^{3/2}$. Here, we don't have such a condition. Therefore we solve the viscous peeling similarity solution numerically without prescribing \textit{a priori} the asymptotics of $H$ at large $X$ (see Appendix~\ref{app:Num_viscous_peeling}). A scaling analysis of the similarity equations reveals the possibility of a similarity solution with $H(X\to \infty) \sim X^{2/3}$, which towards the side of the contact has $H(X\to -\infty) = 1 + \mathcal{O}\left((-X)^{4/3}\right)$, which is manifestly different from the lubricated Hertzian contact~\cite{snoeijer2013similarity}. The resulting similarity solution is superimposed as the dashed line in Fig.~\ref{fig:fig6}(b)-(d). The similarity solution again offers a perfect description of the data and demonstrates that the pressure spike during the snap-off is universal. The only open question is what determines equation~\eqref{eq:snapping-contact_radius}, which remains to be derived in the future. 

\begin{figure*}
	\centerline{\includegraphics{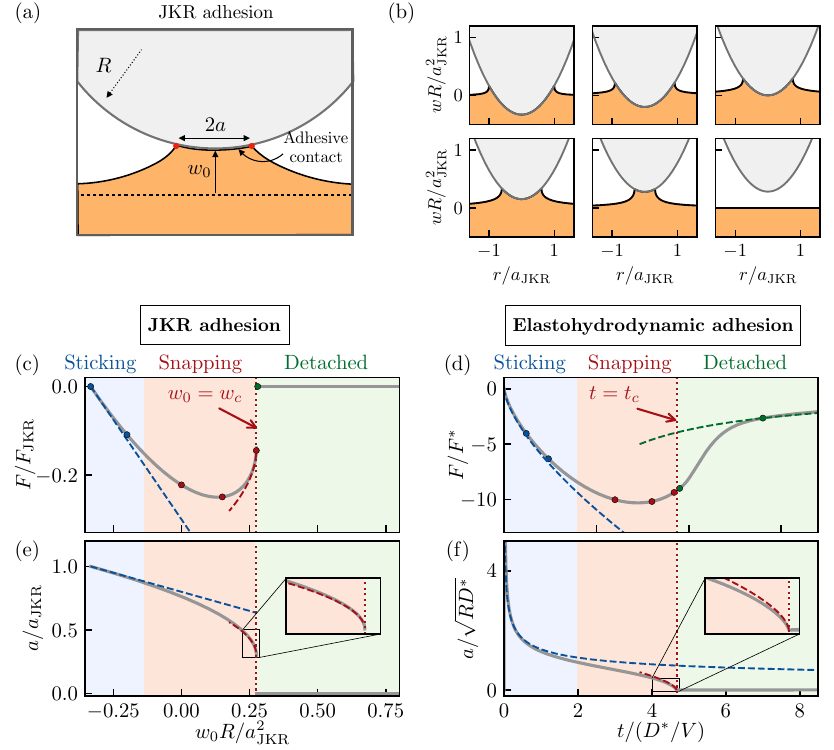}}
	\caption{(a) Schematic of the classical dry adhesion described by JKR theory. In contrast to elastohydrodynamic adhesion, there is a direct contact between the solids, with contact radius $a$ and sphere position $w_0$. (b) Series of snapshots of the elastic displacement fields in JKR model, where $a_\mathrm{JKR}$ is the contact radius at zero force. (c) Normalized force versus normalized sphere vertical position in the JKR model. The normalization force is $F_\mathrm{JKR} = 4/3E^*a_\mathrm{JKR}^3/R = 2\pi \Delta \gamma R$. (e) Normalized contact radius versus normalized sphere vertical position in the JKR model. The red dotted lines indicate the snapping position and the dashed lines the asymptotics near snapping. The blue dashed lines shows the asymptotics at vanishing force. The circles correspond to the snapshots in the panel (b). The panels (d)-(f) show the force and contact radius versus time in elastohydrodynamic adhesion, corresponding to Fig.~\ref{fig:fig3}(c) and Fig.~\ref{fig:fig5}(b) for $h^*/\delta = 10$. The circles indicate the times of the snapshots in Fig.~\ref{fig:fig2}($ii$). The colored background shades highlight the different phases of the dynamics.}
	\label{fig:fig7}
\end{figure*}	

\section{Conclusion and outlook} 
\label{sec:conclusion}
To summarize, we have investigated the viscous adhesive force that occurs when moving a rigid sphere away from a soft surface at a constant speed. The soft surface initially gets deformed following the sphere displacement, such that the sphere effectively sticks to the surface. In this phase, the elastic displacements satisfy a self-similar solution involving both viscosity and elasticity where the effective contact radius decreases in time as $t^{-1/3}$. The adhesive force increases with elastic displacements in a non-linear fashion $F \sim t^{2/3}$, such that the soft surface cannot be approximated by a linear spring with a fixed effective stiffness. When the elastic displacements reach the typical elastohydrodynamic thickness, the solution departs from this self-similar regime and eventually snaps-off violently. The snapping is a finite-time singularity where both the pressure and radial velocity locally diverges, following a universal viscous peeling solution derived here. 

Interestingly, even though there is no direct contact in elastohydrodynamic adhesion, the phenomenology is strikingly similar to the classical JKR adhesion for dry contacts~\cite{johnson1971surface,johnson1987contact}. In the JKR-framework, the sphere directly sticks to a soft surface due to an adhesive surface energy that is liberated upon sticking. Figure~\ref{fig:fig7}(a,b) recalls the classical profiles of the JKR-problem (details in Appendix~\ref{app:JKR}), which can be compared to the profiles in Figure~\ref{fig:fig2}(ii): both the dry and wet adhesion exhibit a sticking phase and a snapping phase. Also the evolution of the force $F$ and contact radius $a$ exhibit similarities, as is quantified in Fig.~\ref{fig:fig7}(c)-(e), comparing the results for JKR (left) and elastohydrodynamic adhesion (right). Most specifically, the snapping occurs at well-defined critical points: critical displacement $w_c$ for JKR and critical time $t_c$ for elastohydrodynamic adhesion. Despite these similarities, some important differences arise. The snapping in JKR occurs at a finite force $F_c$ and finite contact radius $a_c$, resembling a first order transition. The critical point is characterised by scalings $F- F_c \sim a-a_c \propto (w_c-w)^{1/2}$ (see red dashed lines in Fig.~\ref{fig:fig7}(c-e)). By contrast, the elastohydrodynamic snap-off occurs as $a \to 0$, resembling a second order transition, and exhibits a nontrivial exponent $3/5$ (rather than $1/2$). 

Our central finding is thus that soft elastohydrodynamic adhesion is fundamentally different from the Stefan force for rigid surfaces, but exhibits an intricate mechanics that bears similarities with JKR theory. The approach of this article should also apply to other geometries such as cylindrical probes~\cite{nase2008pattern,papangelo2023detachment}, and other soft interfaces like droplets or bubbles during the take-off from surfaces~\cite{boreyko2009self,mouterde2017merging,bashkatov2024performance}. Another interesting extension of this work would be to include viscoelastic properties of the adhesive fluid that play a role for adhesives like the mucus of animals~\cite{noel2017frogs}, or in pressure-sensitive adhesives~\cite{creton2003pressure}. We expect our work to serve a blueprint for sticking and snapping in a broad range of applications involving dynamical adhesion.

\section*{Acknowledgements}  We thank A. Sikkema and E. Wonink  for performing preliminary experiments and P. Chantelot and K. Venner for stimulating discussions. We acknowledge financial support from NWO through the VICI Grant No. 680-47-632. 

\appendix
\section{Boundary conditions of the viscous peeling similarity solution}
\label{app:boundary_conditions}

The governing equations of the viscous peeling similarity solution are equivalent to the ones derived in Ref.~\cite{snoeijer2013similarity}. These authors considered the lubrication problem of an elastic sphere sliding near a planar wall at an imposed speed and normal load. In the small speed (or large load) limit, a Hertzian pressure profile develops within the contact region $p(r<a) =\frac{2}{\pi}E^* \sqrt{a^2-r^2} \sim_{r\to a^-} \frac{2}{\pi}E^*\sqrt{2a} \left(a-r\right)^{1/2}$ and boundary layer arises at the inlet and outlet of the contact. Describing the boundary layer with self-similar solution, the resulting set of equations is equivalent to the one derived here in \eqref{eq:LLD-like} and \eqref{eq:Hilbert}. However, the matching condition to the Hertzian outer problem imposes boundary conditions of the type $P(X\to -\infty) \propto (-X)^{1/2}$ and $H(X\to \infty) \propto X^{3/2} $. 

There is no such requirement in the viscous peeling problem here. We investigate the other possible asymptotic behaviors. We first assume power law asymptotic behavior of the kind $H(X\to\infty) = \alpha X^\beta$, with $\beta>0$. From the equation~\eqref{eq:LLD-like}, the pressure follows the asymptotic behavior $P'(X\to\infty) = (\alpha X^\beta)^{-2} $ and $P(X\to\infty) = \alpha^{-2}/(1-2\beta) \, X^{1-2\beta}$. The Hilbert transform of a power-law is a power-law with the same exponent. Hence, the equation~\eqref{eq:Hilbert} yields to the equation
\begin{equation}
\beta-2 = -2\beta, \quad \rightarrow \quad \beta =2/3.
\end{equation}
The exponent is smaller than unity, which is in qualitative agreement with the concave shape of $H$ observed at the edge of the contact radius.

\section{Numerical resolution of the viscous peeling similarity solution}
\label{app:Num_viscous_peeling}

\begin{figure}
	\centerline{\includegraphics{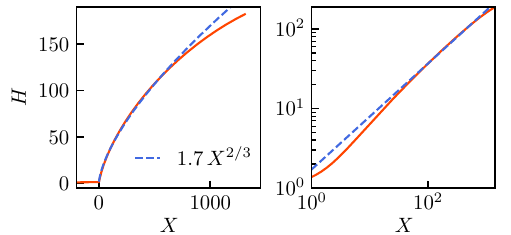}}
	\caption{Numerical similarity solution $H(X)$ of \eqref{eq:LLD-like}-\eqref{eq:Hilbert}, where logarithmic scales are used in the right panel. The dashed line displays the asymptotic behavior at large $X$. Here, the length of the computational domain is $L=1400$. }
	\label{fig:fig8}
\end{figure}	

We solve the coupled system of equations \eqref{eq:LLD-like}-\eqref{eq:Hilbert} numerically with a finite-difference method. To be more specific, we first generate a non-uniform grid $X_i \in [-L, L]$, where $L \gg 1$ is the arbitrary length of the computational domain. The equation is second order such that two boundary conditions are required. To satisfy the film thickness matching condition $H(X\to -\infty) = 1$, we impose a Neumann condition at $X = -L$ (zero slope). At $X=L$, we fix the value of $H$ without any restriction on the shape of $H$. The resulting discrete set of equations is then solved using the root finder algorithm in the scipy.optimize library of Python, which is based on the Powell's method.

The system of equations \eqref{eq:LLD-like}-\eqref{eq:Hilbert} is translational invariant such that the modification of the value of the $H(L)$ translates the solution. An asymptotic behavior of $H(X \to \infty) \simeq \alpha X^{2/3}$ is recovered numerically with $\alpha \approx 1.7$ as shown in Fig.~\ref{fig:fig8}. Deviations from the asymptotic behavior occur around $X \sim 1000$ due to the finiteness (here $L=1400$) of the computational domain. We have checked that the prefactor $\alpha$ is robust to changes in $L$. 

\section{JKR adhesion model}
\label{app:JKR}

The JKR framework assumes that a sphere adheres to a soft surface with an adhesive surface energy $\Delta \gamma$. The vertical position of the sphere and contact radius are denoted $w_0$ and $a$ respectively (see Fig.~\ref{fig:fig7}(a)). Regardless of adhesion, the general solution of the elasticity problem can be expressed as a linear combination of two classical contact mechanics problems, which are the Hertz indentation and the flat cylinder indentation (called flat punch), both of contact radius $a$. The value of the contact radius is found by minimizing the total free energy of the problem, which includes elastic and adhesion energy. The contact radius is related to the sphere vertical position by the relation
\begin{equation}
\label{eq:JKR-displacement}
w_0 = -\frac{a^2}{R} + \frac{2}{3} \, \frac{a^{1/2}a_\mathrm{JKR}^{3/2}}{R}, \quad a_\mathrm{JKR}^3 = \frac{3 \pi \Delta \gamma R^2}{2E^*},
\end{equation}
where we introduce $a_\mathrm{JKR}$ the contact radius at vanishing force. Similarly the force $F$ can be expressed as function of the contact radius $a$ as
\begin{equation}
\label{eq:JKR-force}
F = \frac{4E^*}{3R}\left(a^3 - a^{3/2}a_\mathrm{JKR}^{3/2}\right),
\end{equation}
where the first term corresponds to the usual Hertz indentation, while the second one describes adhesion effects. The elastic displacement fields is a piecewise-defined function that reads
\begin{subequations}
\label{eq:JKR_displacement_field}
\begin{equation}
\begin{split}
w(r) = &-\frac{a^2}{\pi R}\left[ \left(2-\frac{r^2}{a^2}\right)\,\text{arcsin}\frac{a}{r}+ \sqrt{\frac{r^2}{a^2}-1} \right]\\
& + \frac{4a^{1/2}a^{3/2}_\mathrm{JKR}}{3\pi R}\, \text{arcsin}\frac{a}{r} \, \quad \text{for} \quad r>a,
\end{split}
\end{equation}
\begin{equation}
w(r) = w_0 + \frac{r^2}{2R} \, \quad \text{for} \quad r<a.
\end{equation}
\end{subequations}

\section{Asymptotic analysis of JKR model}

In this subsection, we perform the asymptotic expansion of the JKR model of section~\ref{app:JKR} around the two points of interest, which are the zero force $F = 0$ and the snapping point $w_0 = w_c$. 

\subsection{Expansion at zero force: spring-like behavior}  
\label{app:JKR_F=0}
We start with the behavior for vanishing force, which corresponds to a vertical position of the sphere $w_0 = -\frac{1}{3} \frac{a_\mathrm{JKR}^2}{R}$. We expand the vertical position as $w_0 = -\frac{1}{3} \frac{a_\mathrm{JKR}^2}{R} + \delta w$ and inject it into \eqref{eq:JKR-displacement}-\eqref{eq:JKR-force} to find the leading-order expansion as
\begin{subequations}
\begin{equation}
\label{eq:JKR-a_spring}
a  = a_\mathrm{JKR} - \frac{3}{5} \frac{R}{a_\mathrm{JKR}}  \delta w = a_\mathrm{JKR} - \frac{3R}{5a_\mathrm{JKR}} \left(w_0+ \frac{a_\mathrm{JKR}^2}{3R} \right),
\end{equation}
\begin{equation}
\label{eq:JKR-Force_spring}
F = -\frac{6E^* a_\mathrm{JKR}}{5}  \delta w = -\frac{6E^* a_\mathrm{JKR}}{5} \left(w_0+ \frac{a_\mathrm{JKR}^2}{3R} \right).
\end{equation}
\end{subequations}
Hence, the adhesion force is linear in sphere vertical position for vanishing load, behaving as an effective linear spring. The contact radius also decreases linearly with increasing sphere position. 

\subsection{Expansion near snapping: saddle-node bifurcation}  
\label{app:JKR_w=wc}
The JKR equations~\eqref{eq:JKR-displacement} only admit solutions up to a maximum sphere vertical position $w_0 < w_c$, where $w_c$ corresponds to the snapping point. Indeed, the curve $a(w_0)$ has an infinite slope $\mathrm{d}a/\mathrm{d}w_0 \to \infty$ for $a = a_c = \frac{1}{6^{2/3}}\, a_\mathrm{JKR}$, corresponding to sphere vertical position $w_c = \frac{1}{2}\frac{1}{ 6^{1/3}}  \frac{a_\mathrm{JKR}^2}{R}$. We again expand the sphere vertical position as $w_0 = w_c - \delta w$ and find the behavior near snapping as
\begin{equation}
\label{eq:JKR-a_snapping}
a = a_c +\sqrt{\frac{2R}{3}} \left(w_c - w_0\right)^{1/2},
\end{equation}
\begin{equation}
\label{eq:JKR-Force_snapping}
\begin{split}
F =  F_c &-\frac{4 E^* a_\mathrm{JKR}^2}{3 \cdot 6^{1/3}R}\sqrt{\frac{2R}{3}} \left(w_c - w_0\right)^{1/2}, \\ & \text{with} \quad F_c = -\frac{5E^* a_\mathrm{JKR}^3}{27 R}.
\end{split}
\end{equation}
We notice the non-linear equation \eqref{eq:JKR-displacement} for $a(w_0)$ admits an another branch for $a<a_c$ that is not represented in Fig.~\ref{fig:fig7}(e) for the sake of clarity. Indeed, this branch is unstable, and the location of the snapping point correspond to the point where both branches meet. Hence, the snapping in JKR theory can be classified as a saddle-node bifurcation.

\providecommand{\noopsort}[1]{}\providecommand{\singleletter}[1]{#1}%

\end{document}